\documentclass[preprint,aps,12pt,preprintnumbers,eqsecnum,nofootinbib]{revtex4}
\usepackage{graphicx}
\usepackage{subfigure}

\usepackage{color}
\usepackage{amssymb,amsmath,amsfonts}
\usepackage{epstopdf} 
\usepackage{braket}

\newcommand{\beq}{\begin{equation}}
\newcommand{\enq}{\end{equation}}

\newcommand{\bk}{\braket}
\newcommand{\x}{\times}

\newcommand{\cd}{\cdot}
\newcommand{\da}{\dagger}

\newcommand{\pa}{\partial}
\DeclareMathOperator{\tr}{tr}

\usepackage{slashed}

\newcommand{\xone}{\chi^{(1)}}
\newcommand{\xtwo}{\chi^{(2)}}
 
\newcommand{\M}{\mathcal{M}}
\newcommand{\lara}[1]{\left\langle #1 \right\rangle}

\begin{document}
%
%
\title{\vspace*{0.5in} 
Dark sector portal with vector-like leptons and flavor sequestering
\vskip 0.1in}
\author{Christopher D. Carone}\email[]{cdcaro@wm.edu}
\author{Shikha Chaurasia}\email[]{scchaurasia@email.wm.edu}
\author{Tangereen V. B. Claringbold}\email[]{tvclaringbold@email.wm.edu}

\affiliation{High Energy Theory Group, Department of Physics,
College of William and Mary, Williamsburg, VA 23187-8795}

\date{\today}

\begin{abstract}
We consider models with fermionic dark matter that transforms under a non-Abelian dark gauge group.  Exotic, vector-like
leptons that also transform under the dark gauge group can mix with standard model leptons after spontaneous symmetry breaking and 
serve as a portal between the dark and visible sectors.  We show in an explicit, renormalizable model based on a dark SU(2) gauge group 
how this can lead to adequate dark matter annihilation to a standard model lepton flavor so that the correct relic density is obtained.  We 
identify a discrete symmetry that allows mass mixing between the vector-like fermions and a single standard model lepton flavor, while 
preventing mixing between these fields and the remaining standard model leptons.  This flavor sequestering avoids unwanted 
lepton-flavor-violating effects, substantially relaxing constraints on the mass scale of the vector-like states.  We discuss aspects of the phenomenology of 
the model, including direct detection of the dark matter.
\end{abstract}
\maketitle

\section{Introduction}
Although the literature on dark matter models is vast and diverse, the organizational structure of many of these models 
is similar.   The visible sector includes all the fields normally associated with the minimal standard model;  the dark sector 
consists of a collection of fields that communicate very weakly with the visible sector; the messenger or portal sector
consists of those fields that allow for a weak coupling between the visible and dark sectors.   In this paper, we are 
interested in a possible portal for non-Abelian dark matter models, specifically ones in which fermionic dark matter is 
charged under a non-Abelian dark gauge group.    Examples of non-Abelian dark matter models can be found in 
Refs.~\cite{Choquette:2015mca,Cheung:2014tha,Cline:2014kaa,Cline:2010kv,Carone:2010ha,Chen:2009ab,Zhang:2009dd} 
and~\cite{Davoudiasl:2013jma,Chiang:2013kqa},  though we will not focus on models like those in 
Refs.~\cite{Davoudiasl:2013jma,Chiang:2013kqa} where the dark gauge boson is itself the dark matter.   We are interested 
here in a mechanism that allows the dark gauge boson to develop a small coupling to the visible sector, adequate enough to 
facilitate the annihilation of the dark matter for a successful thermal freeze-out, without running afoul of direct detection bounds.  
We focus on a model in which the effective coupling between the dark and visible sectors does not appear at the same order in 
the dark matter annihilation and the dark-matter-nucleon elastic scattering cross sections.

One approach is to arrange for couplings between the non-Abelian dark gauge bosons and standard model leptons, but 
not quarks; in this case, dark matter annihilation would proceed via tree-level diagrams, while diagrams involving quarks 
would be higher-order.  However, this simple approach can lead to significant model building complications.   For 
example,  if one tries to couple the dark gauge bosons to the standard model leptons directly, then the dark gauge 
bosons are potentially no longer ``dark,'' unless their gauge coupling is taken to be very small.   However, this choice 
suppresses the coupling of the dark gauge bosons to both the dark and visible sectors, making it ineffective as a 
channel for dark matter annihilation.  Moreover, such direct couplings lead generically to chiral 
anomalies, which must be cancelled by additional states that are charged under both the dark and standard model gauge 
groups.    There is no guarantee that the simplest Higgs field content of the dark and visible sectors will have the correct 
quantum numbers to provide Yukawa couplings for these additional states, so that 
additional Higgs representations may be required.   Another potential problem is that charging standard model leptons 
under the new non-Abelian group may either restrict the form of the standard model lepton Yukawa matrices in unwanted
ways, or forbid them entirely, unless a Higgs field charged under both the dark and standard model gauge groups is introduced.  
While the proliferation of fields implied by these considerations does not rise to the level of a no-go theorem, it does make the approach described a lot less appealing.
  
Fortunately, there is a simple way to avoid the complications described above:  the non-Abelian dark gauge boson may 
couple to a vector-like state $\chi$ that can mix with standard model leptons {\em after} the gauge symmetries of the 
theory (both dark and visible) are spontaneously broken.  We will refer to the $\chi$ states as heavy, vector-like leptons.   
If the dark gauge boson's coupling to dark matter is $g_D$, which may be substantial, then the induced coupling to the 
standard model lepton in the mass eigenstate basis will be proportional to $\theta g_D$ where $\theta$ is a small mixing angle. Since 
the gauge boson couples directly to a vector-like state, anomalies are cancelled, and a mass term 
$-M_\chi \overline{\chi} \, \chi$ can be written down at tree-level.  The range of $M_\chi$ is then determined by 
phenomenological requirements, for example, that the mixing angle $\theta$ is large enough to produce the desired value 
of the dark matter relic density via dark matter annihilation to a standard model lepton-anti-lepton pair.  In this paper, 
we will present an explicit and renormalizable model that illustrates this proposal.  Our focus differs from that of
Refs.~\cite{Choquette:2015mca,Cheung:2014tha,Cline:2014kaa,Cline:2010kv,Carone:2010ha,Chen:2009ab,Zhang:2009dd}, where 
the origin of higher-dimension operators connecting the dark and visible sectors was either unspecified, or assumed to arise from a 
sector whose flavor structure and phenomenology was not explicitly investigated.

To obtain a sufficient dark matter annihilation cross section to standard model particles, the mixing angle $\theta$ cannot 
be too small; this implies that the heavy, vector-like leptons $\chi$ cannot be arbitrarily heavy.   As we will see later, a 
$\chi$ mass of order $100$~TeV would not naturally lead to the desired annihilation cross section.  At face value, this seems 
problematic, since the literature includes bounds on vector-like heavy leptons that exceed $100$~TeV~\cite{Ishiwata:2013gma}.  Such 
stringent bounds, however, come from consideration of lepton-flavor-violating processes that emerge when the vector-like 
states mix with all three standard model lepton flavors.   One expects such mixing to be present generically, and this 
would doom the approach that we have just outlined.   In this paper, we show how a more favorable outcome can be 
achieved via discrete symmetries that allow us to suppress the unwanted mass mixing arbitrarily.  In our model, 
vector-like leptons mix only with a single flavor of the standard model leptons, which in turn 
does not mix substantially with the remaining two flavors, thus avoiding problems with lepton flavor violation.   We refer to
this restricted pattern of mass mixing as ``flavor sequestering."  We will show that the discrete symmetry used to achieve this result does 
not adversely affect the remaining flavor structure of the charged leptons or neutrino mass matrices.   
Phenomenological considerations place constraints on the mass spectrum of the flavor-sequestered vector-like lepton 
states that can be tested in direct collider searches.

Our paper is organized as follows.   In the next section, we define the simplest model that illustrates a portal involving 
vector-like leptons and flavor sequestering.  In Sec.~3, we show how the flavor structure of the theory can be achieved using 
a discrete symmetry, so that exclusive mixing with one standard model lepton generation is obtained and lepton-flavor-
violating effects avoided.  In Sec.~4 we discuss phenomenological constraints on the model parameters, including the 
requirement that the correct dark matter relic density is obtained, and the constraints from dark matter-nucleon elastic 
scattering, which follows from kinetic mixing that is induced after the non-Abelian gauge group is spontaneously broken. 
In the final section, we summarize our conclusions.

\section{The Model} \label{sec:model}

We consider the simplest non-Abelian dark gauge group, SU(2)$_D$.   As stated earlier, we denote the heavy, vector-like leptons 
$\chi$, and assume the quantum numbers
\begin{equation}
\chi_L \sim \chi_R \sim ({\bf 2},{\bf 1},{\bf 1},-1) \,\,\, ,
\end{equation}
where we indicate the representations of SU(2)$_D \times$ SU(3)$_C\times$ SU(2)$_W\times$U(1)$_Y$, in that order.  In other words, these states 
are SU(2)$_D$ doublets, but have the same electroweak charges as right-handed leptons.   We further assume the simplest assignment for the 
dark matter, {\em i.e.}, that it is a doublet under SU(2)$_D$.  However, to avoid a Witten anomaly~\cite{Witten:1982fp} there must be an even number of 
SU(2) fermion doublets, 
so we take 
\begin{equation}
\psi_L \sim \psi_R \sim ({\bf 2},{\bf 1},{\bf 1},0)\,\,\, .
\end{equation}
Since the $\psi$ fields are charged only under SU(2)$_D$, we can construct Dirac or Majorana mass terms, or both.  We will assume
Dirac mass terms, for simplicity, though it is easy to make this the only possibility by imposing additional discrete symmetries.  For 
example, an unbroken $Z_3$ symmetry can forbid Majorana masses for $\psi$, and also serve as the symmetry which stabilizes the dark 
matter, which we identify henceforth as the lightest component of the $\psi$ doublet.

We assume that the dark gauge symmetry is spontaneously broken by two SU(2)$_D$ Higgs field representations,
\begin{equation}
H_D \sim ({\bf 2},{\bf 1},{\bf 1},0) \,\,\,\,\, \mbox{  and  } \,\,\,\,\, H_T \sim ({\bf 3},{\bf 1},{\bf 1},0) \,\,\,\, .
\end{equation}
We show at the end of this section that the Higgs potential has local minima consistent with the pattern of vacuum expectation values (vevs):
\begin{equation}
\langle H_D \rangle = \left(\begin{array}{c} v_{D1} \\ v_{D2} \end{array}\right) \,\,\,\,\, \mbox{ and } \,\,\,\,\, \langle H_T \rangle = \left(\begin{array}{cc}
v_T/2 & 0 \\ 0 & -v_T/2 \end{array}\right) \,\,\, . \label{eq:hdhtv}
\end{equation}
If we decompose $H_T = H_T^a \,(\sigma^a/2)$, where the $\sigma^a$ are Pauli matrices, then the $H_T$ vev above corresponds to $\langle H_T^3 \rangle = v_T$ and
$\langle H_T^a \rangle = 0$ for $a=1, 2$.   In fact, an arbitrary vev for $H_T$ can always be rotated into the $H_T^3$ direction by an SU(2)$_D$ transformation.   
With this choice, vevs in both components of $H_D$ are expected, and one of those can be made real by a further SU(2)$_D$ phase rotation.   The fact that the remaining 
$H_D$ vev in Eq.~(\ref{eq:hdhtv}) is assumed real will be shown to be consistent with the minimization of a potential later.

\begin{figure}[t]
  \begin{center}
    \includegraphics[width=0.3\textwidth]{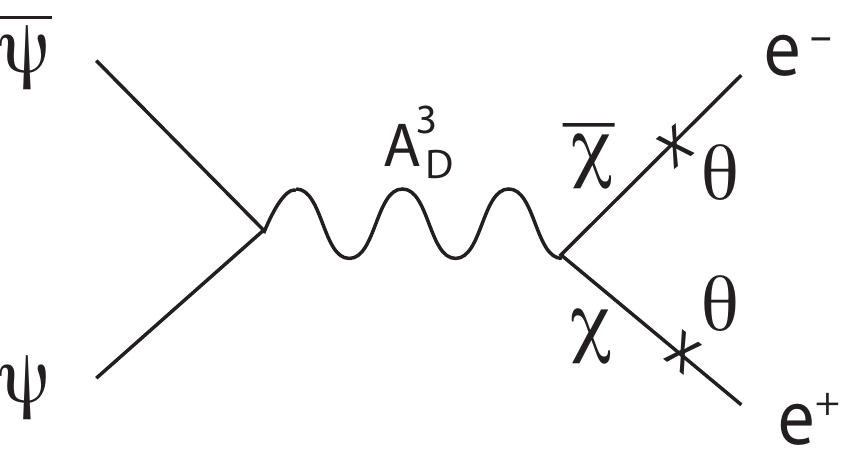}
    \caption{ Qualitative picture of dark matter annihilation to a charged lepton-anti-lepton pair, as discussed in the text. The insertions labelled by $\theta$ represent mass mixing. \label{fig:one}}   
      \end{center}
\end{figure}

We can now say something more concrete about the mass spectrum of the model.  The relevant Lagrangian terms are ${\cal L} \supset {\cal L}_\psi + {\cal L}_{\chi e}$, where
\begin{equation}
{\cal L}_\psi = - M_\psi\, \overline{\psi}_L\, \psi_R + \lambda_s \, \overline{\psi}_L\, H_T \,\psi_R + \mbox{ h.c. }  \,\,\,\, , \label{eq:lpsi}
\end{equation}
and
\begin{equation}
{\cal L}_{\chi e} = - M_\chi\, \overline{\chi}_L\, \chi_R + \lambda_s' \, \overline{\chi}_L\, H_T \,\chi_R - y_1 \, \overline{\chi}_L \,H_D\, e_R 
- y_2  \, \overline{\chi}_L \,\widetilde{H}_D\, e_R- y_e \,  \overline{L}_L\, H \,e_R + \mbox{ h.c. }   \,\,\, ,  \label{eq:lchi}
\end{equation}
where $\widetilde{H}_D \equiv i \sigma^2 H_D^*$, and the final term is the usual standard model Yukawa coupling for a single lepton flavor.  Eq.~(\ref{eq:lchi}) assumes 
the existence of a symmetry that leads to exclusive mixing between any one standard model, right-handed charged lepton flavor (called $e_R$ above) and the 
vector-like $\chi$ fields.  We show how this flavor sequestering can be arranged by a discrete symmetry in Sec.~\ref{sec:zn}.    The first terms in Eqs.~(\ref{eq:lpsi}) and (\ref{eq:lchi}) provide a common mass for each component of the given doublet,
while the second terms lead to mass splittings proportional to the vev $v_T$. The third 
and fourth terms in Eq.~(\ref{eq:lchi}) allow mixing between the standard model lepton $e_R$ and the $\chi$ fields, since the coupling 
to the dark doublet Higgs field $H_D$ allows for the formation of an SU(2)$_D$ singlet.  The final term leads to an $e$ mass when the standard model Higgs field develops a 
vacuum expectation value $\langle H \rangle = (0 , v/\sqrt{2})$, with $v = 246$~GeV.  Defining the column vector  $\Upsilon \equiv (e, \,\chi^{(1)},\, \chi^{(2)})^T$, which displays the two components of the $\chi$ doublet,  we may write the mass matrix that is produced after spontaneous symmetry breaking by
\begin{equation}
{\cal L}_{mass}^{\chi e} = - \overline{\Upsilon}_L  M \,\Upsilon_R + \mbox{ h.c. }  \,\,\, ,
\end{equation}
where
\begin{equation}
M = \left(\begin{array}{ccc}  \frac{h_e v}{\sqrt{2}}  & 0 & 0 \\
\frac{(y_1 v_{1D} + y_2 v_{2D})}{\sqrt{2}} & M_\chi - \frac{\lambda_s'v_T}{2} & 0 \\
\frac{(y_1 v_{2D} - y_2 v_{1D})}{\sqrt{2} }   & 0 &  M_\chi + \frac{\lambda_s'v_T}{2}  \end{array}\right) \equiv \left(\begin{array}{ccc} m_0 & 0 & 0 \\ m_1 & M_1& 0 \\ m_2 & 0 & M_2
\end{array}\right) \,\,\, , \label{eq:bigM}
\end{equation}
where the second form is a convenient parametrization.  This matrix can be diagonalized by a bi-unitary transformation,  $M = U_L \, M^{diag} \, U_R^\dagger$.   While this diagonalization can be done numerically, there are certain limits that are relevant to us in which simple 
results can be obtained.   In particular, when  $M_1$,  $M_2$ $>>$  $m_1$, $m_2$ $>>m_0$, we find that the largest mixing angles, which occur in $U_R$, are given by
\begin{equation}
U_R = \left(\begin{array}{ccc}  1 -\frac{1}{2}\left(\frac{m_1^2}{M_1^2}+\frac{m_2^2}{M_2^2}\right)
& m_1/M_1 & m_2/M_2 \\  -m_1/M_1 & 1-\frac{1}{2} \frac{m_1^2}{M_1^2}  & -\frac{M_1}{M_2}\frac{m_1 m_2}{M_1^2-M_2^2} \\ -m_2/M_2 
& \frac{M_2}{M_1}\frac{m_1 m_2}{M_1^2-M_2^2} & 1-\frac{1}{2} \frac{m_2^2}{M_2^2} \end{array}\right) + \cdots \,\,\, , \label{eq:ur}
\end{equation}
where the $\cdots$ represent terms that are cubic order or higher in $m_i/M_j$.   For this case, we can now find the leading coupling of the dark gauge fields $A_{D\mu}^a$ to the mass eigenstate fields.  In the gauge basis, the coupling to $\Upsilon_R$ can be written
\begin{equation}
{\cal L} = i \overline{\Upsilon}_R  \gamma^\mu (\partial_\mu - i g_D A^a_{D\mu} {\cal T}^a) \Upsilon_R + \cdots  \,\,\, ,
\end{equation}
where  
\begin{equation}
{\cal T}^a = \left(\begin{array}{c|c} 0 & 0\\  \hline 
0 & \, T^a \end{array} \right)  \,\,\, ,
\end{equation}
and $T^a = \sigma^a/2$, $a=1, \ldots, 3$,  are the generators of SU(2).  The zero in the $1$-$1$ element reflects the fact that the standard model lepton is not charged
under the dark gauge group.  In the mass eigenstate basis, the couplings of the $a^{th}$ dark gauge boson are therefore proportional to $U_R^\dagger {\cal T}^a U_R$.  In the same approximation as Eq.~(\ref{eq:ur}), these matrices are given by
\begin{eqnarray}
&&U_R^\dagger {\cal T}^a U_R = \nonumber \\ 
  &&\left[ \left(\begin{array}{ccc} \frac{m_1 \, m_2}{M_1 \, M_2} & -\frac{m_2}{2 \, M_2} & -\frac{m_1}{2\, M_1} \\
 -\frac{m_2}{2 \, M_2} & 0 & \frac{1}{2} \\
 -\frac{m_1}{2\, M_1} & \frac{1}{2} & 0 \end{array} \right) \,\,\, ,\,\,\,
 \left(\begin{array}{ccc} 0 & -\frac{i \, m_2}{2 \, M_2} & \frac{i \, m_1}{2\, M_1} \\
 \frac{i \, m_2}{2 \, M_2} & 0 & -\frac{i}{2} \\
 -\frac{i\, m_1}{2\, M_1} & \frac{i}{2} & 0 \end{array} \right)  \,\,\, , \,\,\,
 \left(\begin{array}{ccc} \frac{m_1^2}{2\, M_1^2} - \frac{m_2^2}{2\, M_2^2} & -\frac{m_1}{2 \, M_1} & \frac{m_2}{2\, M_2} \\
 -\frac{m_1}{2 \, M_1} & \frac{1}{2} & 0 \\
 \frac{m_2}{2\, M_2} & 0 & -\frac{1}{2} \end{array} \right) \right] \,\,\,, \nonumber \\ \label{eq:utu}
 \end{eqnarray}
where we only show results to linear order in $m_i/M_j$, with the exception of the $1$-$1$ entries, because of their relevance to our subsequent
discussion.  For example, for the lightest dark gauge boson, $A^3_D$, the coupling to  $e^+\, e^-$ is given by
\begin{equation}
g_D \overline{\Upsilon}_R  \gamma^\mu  A^3_{D\mu} (U_R^\dagger{\cal T}^3 U_R) \Upsilon_R  = \frac{g_D}{2} \left(\frac{m_1^2}{M_1^2}-\frac{m_2^2}{M_2^2} \right) \overline{e}_R \gamma^\mu  A^3_{D\mu} e_R + \cdots
\end{equation}
which provides the $A_D^3$ gauge boson with a decay channel (since we assume its mass is greater that $2 \,m_e$) and provides the dominant portal for dark matter 
annihilation into standard model particles.   For later convenience, we define
\begin{equation}
\theta^2 \equiv g_D\left(\frac{m_1^2}{M_1^2}-\frac{m_2^2}{M_2^2} \right) \,\,\, . \label{eq:ourtheta}
\end{equation}
We illustrate the qualitative idea in Fig.~\ref{fig:one}  that the dark matter annihilation process of interest emerges from mixing that affects
two of the external legs.

We note that in the case where $m_0$ is comparable to $m_1$ and $m_2$ we find via numerical diagonalization that our expression $U_R$
in Eq.~(\ref{eq:ur}) still provides an accurate approximation.  Moreover, we can prove that $m_0$ appears only as a higher-order correction to
$\theta$, as defined in Eq.~(\ref{eq:ourtheta}), the quantity that is most relevant to our phenomenological discussion later.   The argument is as follows:  if $m_1$ or $m_2$ where to vanish, then $U_R$ must become the identity.  This implies that any corrections to the 
$1$-$2$, $1$-$3$, $2$-$1$ and $3$-$1$ entries of $U_R$ that are proportional to $m_0$ must come at no lower order than $m_0 m_{1,2} / M_{1,2}^2$. This potential contribution is nonetheless higher-order than the values  shown for these entries in Eq.~(\ref{eq:ur}).  It is also the case that the $1$-$1$ entry of $U_R^\dagger {\cal T}^3 U_R$, from which $\theta$ is extracted, depends only on these four entries.  Hence,
the value of $\theta$, which controls the induced coupling of $A^3_D$ to the chosen standard model lepton flavor, remains unaffected at leading order.

Eq.~(\ref{eq:utu}) indicates that all states other than the lightest $\psi$ mass eigenstate have available decay channels that ultimately lead to standard model particles.  Let us give 
some examples to establish this point, assuming the dark sector mass scale is significantly larger than the electron mass scale (we assume $>2$~GeV in Sec.~\ref{sec:pheno}), and that the 
dark gauge bosons are heavier than the dark matter fermion.   The coupling matrices $U_R^\dagger {\cal T}^a U_R$, for $a=1$ and $a=3$ allow decays of $A^1_D$ and $A_D^3$ directly 
to $e^+\, e^-$; the same is not true for $a=2$, but the $A^2_D$ boson does couple to the two different $\psi$ mass eigenstates, which we will call $\psi^{(1)}$ (the lighter, dark matter component) 
and $\psi^{(2)}$ (the heavier).  The eigenstate $\psi^{(2)}$ can decay to dark matter $\psi^{(1)}$ plus $e^+ \, e^-$ via $A^1_D$ exchange.   Hence $A^2_D$ can decay to two dark matter particles 
and an $e^+ \, e^-$ pair, whether or not $\psi^{(1)}$ is on shell.  Due to the $\chi A_D^{3} e$ couplings in $U_R^\dagger {\cal T}^3 U_R$,  both $\chi$ mass eigenstates can decay to a 
same-sign $e$ plus an $e^+\, e^-$ pair via $A_D^3$ exchange. Finally, the exotic Higgs fields $H_D$ and $H_T$ couple to fermion pairs via their Yukawa couplings.  Since we have already established that those fermions couple ultimately to either $e$'s or $\psi^{(1)}$'s, our claim is established.   

Since the $\chi$ and $e_R$ have identical electroweak quantum numbers, there is no effect on the coupling of the $Z$ boson to $e_R$ in the mass eigenstate basis.  However, $\chi$ 
and $e_L$ couple differently to the electroweak gauge bosons, and diagonalization of Eq.~(\ref{eq:bigM}) also involves a left-handed rotation matrix $U_L$ which differs from the identity.   
Fortunately, the left-handed mixing angles are much smaller than those in Eq.~(\ref{eq:ur}) so that this does not present any phenomenological difficulties.  For example, the fractional shift in 
the standard model $Z e_L e_L$ vertex is of ${\cal O}(\frac{m_0 m_1}{M_1^2}\frac{m_0 m_2}{M_2^2})$, which is negligible given the spectrum we assume in Sec.~\ref{sec:pheno}.  We also may take the 
mostly $\chi$ mass eigenstates to be heavy enough so that rare $Z$ decays to $\chi \, e$ are kinematically forbidden.

Finally, let us return to the issue of the spontaneous breaking of the dark gauge symmetry.  In the effective theory well below the electroweak scale, the most general renormalizable potential 
involving the dark Higgs fields is given by
\beq
\begin{split}
\label{eq:higgspotential}
V(H_D, H_T) &= -m_D^2 H_D^\da H_D -m_T^2 \tr(H_T H_T) 
+\lambda_1(H_D^\da H_D)^2 
+ \lambda_2 \big[\tr(H_T H_T)\big]^2 
\\ &+ \lambda_3 H_D^\da H_T H_T H_D 
+\mu_1 H_D^\da H_T H_D+ \left(\mu_2 H_D^\da H_T \widetilde H_D + \text{h.c.}\right),
\end{split}
\enq
where we have used the fact that $H_T^\da = H_T.$ We assume the potential does not violate CP, so that all the couplings are real. Further, we require at least one of 
($-m_D^2$, $-m_T^2$) to be negative so that the $H_D$ and $H_T$ fields may develop non-zero vevs. It should be noted that there are other terms involving the Higgs fields that could be added to the potential, such as $\widetilde H_D^\da \widetilde H_D, \, \tr( H_T^4), \, 
\widetilde H_D^\da H_T \widetilde H_D, \, H_D^\da H_D \tr(H_T H_T)$, but these are not linearly independent of the terms included in Eq.~(\ref{eq:higgspotential}) and so have been omitted. 

The Higgs doublet assumes the standard real-field parametrization,
\beq
\label{eq:hd}
H_D = \frac{1}{\sqrt2}\begin{pmatrix} \phi_1 + i\phi_2 \\ \phi_3 + i\phi_4\end{pmatrix},
\enq
while the Higgs triplet can be represented by a $2\x2$ matrix of real fields $H_1, H_2$ and $H_3$,
\beq
\label{eq:ht}
H_T = H^a \frac{\sigma^a}{2} = \frac{1}{2}\begin{pmatrix} H_3 & H_1 - iH_2 \\ H_1+iH_2 & -H_3 \end{pmatrix}.
\enq
The normalization assures canonical kinetic terms. We proceed to show that there exists a stable, local minimum of the potential for the pattern of vacuum expectation values described in Eq.~(\ref{eq:hdhtv}). One approach to studying the potential is to fix all the parameters and search for minima, using standard steepest descent algorithms. However the downside to this approach is that one may then have to repeatedly discard local minima that do not provide the pattern of vevs desired for the model. So instead, we will fix the vevs and work backwards, showing that an extremum exists that is also a local minimum for a fixed set of parameters.  

The extremization of Eq.~(\ref{eq:higgspotential}) with the fields set to the vevs shown in Eq.~(\ref{eq:hdhtv})
provides the following nontrivial, linearly independent constraints:
\beq 
\begin{split}
-m_D^2 v_{D1}+\lambda_1 v_{D1}^3+\lambda_1 v_{D1} v_{D2}^2+\mu_2 v_{D2} v_T+\frac{1}{4} v_{D1} v_T (\lambda_3 v_T+2 \mu_1) &= 0
\\ \frac{1}{2} \left(-\mu_2 v_{D1}^2+\mu_1 v_{D1} v_{D2}+\mu_2 v_{D2}^2\right) &= 0
\\ -m_T^2 v_T+\frac{1}{4} \lambda_3 v_T \left(v_{D1}^2+v_{D2}^2\right)+\frac{1}{4} \mu_1 (v_{D1}^2-v_{D2}^2) +\mu_2 v_{D1} v_{D2}+\lambda_2 v_T^3 &= 0.
\end{split}
\enq
For the purpose of numerical evaluation we work here in units where $\mu_1=1$. For fixed choices of the vevs and the couplings 
$\lambda_{1,2,3}$, we may then determine $m_D, \, m_T$ and $\mu_2$.  To determine whether the extremum is a minimum, maximum 
or saddle point, we need to examine the eigenvalues of the mass squared matrix (the second derivative matrix with all fields set to their 
vevs and with the solutions for $m_D$, $m_T$ and $\mu_2$ corresponding to the extremum).  Since $SU(2)_D$ is spontaneously broken 
to nothing, we expect three Goldstone bosons, one for each broken SU(2) generator, according to Goldstone's theorem.  Thus we would 
expect three of the eigenvalues to be zero, corresponding to the massless degrees of freedom that are ``eaten" by the dark gauge bosons. 
The remaining eigenvalues must be positive for the extremum to be a local minimum. For example, let us set 
$v_T=v_{D1} = v_{D2}/2 = \lambda_{1,2,3}= \mu_1$ (here we require $v_{D1} \neq v_{D2}$ for a solution to exist). Then we find $m_D^2 = 53/12, \, m_T^2 = 1/6$ and $\mu_2 = -2/3$. The corresponding mass squared eigenvalues are  
$\{0, \, 0, \, 0, \, 3.75, \, 3.75, \, 4, \, 10\},$ in units of $\mu_1^2$, thus confirming that we are at a local minimum of the potential.   This provides an existence proof that local minima exist in which the pattern of vevs shown in Eq.~(\ref{eq:hdhtv}) is obtained.  It is not difficult to find similar solutions for other choices of $v_{D1}$, $v_{D2}$ and $v_T$.

The SU(2)$_D$ breaking vevs affect the $\chi$-$e$ mass spectrum via Eq.~(\ref{eq:bigM}); the triplet vev also splits the $\psi$ mass
eigenstates
\beq
m_{\psi^{(1)}} = M_\psi - \frac{1}{2}\lambda_s v_T, \quad m_{\psi^{(2)}} = M_\psi + \frac{1}{2}\lambda_s v_T
\enq
for
$\psi_{L,R} = \left( \psi^{(1)}, \; \psi^{(2)}\right)_{L,R}^T$.  The gauge field spectrum is obtained from the kinetic terms for $H_D$ and $H_T$,
\beq
 \mathcal L_{kin}(H_D, H_T) = (D_\mu H_D)^\da (D^\mu H_D) + \tr \big[ (D_\mu H_T)^\da (D^\mu H_T)\big],
\enq
where $D_\mu H_D = \pa_\mu H_D - ig_D A_{D\mu}^a \tfrac{\sigma^a}{2} H_D$ 
and $D_\mu H_T = \pa_\mu H_T - i g_D \tfrac{\sigma^a}{2} A_{D\mu}^a H_T + ig_D A_{D\mu}^a H_T \tfrac{\sigma^a}{2}$. Following symmetry breaking the gauge bosons develop masses
\beq
 m_{A_D^1}^2 = m_{A_D^2}^2 = \frac{g_D^2}{4}(v_{D1}^2+v_{D_2}^2+4\, v_T^2), \quad m_{A_D^3}^2 = \frac{g_D^2}{4}(v_{D1}^2+v_{D_2}^2). 
\enq
In splitting the $\psi$ and $A_D$ multiplet masses, the triplet vev leads to a simple low-energy effective theory consisting of the dark matter 
$\psi^{(1)}$ (we assume $\lambda_s>0$) and the mediator $A_D^3$, which has small induced couplings to a right-handed standard model lepton flavor.  This effective theory is relevant below the masses of the heavy vector-like leptons, $\psi^{(2)}$  and the $A_D^{1,2}$ bosons, which we will associate with a common scale, for simplicity.  In addition, we will see that the triplet vev leads to induced couplings of the
dark matter to quarks via kinetic mixing, which will lead to avenues for direct detection.   We discuss the phenomenology of this scenario in Sec.~\ref{sec:pheno}.   

\section{Flavor sequestering} \label{sec:zn}

In this section, we show that it is possible to allow for non-negligible mixing between one flavor of the standard model leptons and the heavier,  
vector-like leptons, while suppressing the mixing with the other standard-model flavors, so that bounds on lepton-flavor-violating processes 
become irrelevant.   In the discussion below, we refer to that one flavor as the electron $e$, though the approach described applies equally 
well if the chosen flavor were $\mu$ or $\tau$. Let us consider the structure of the standard model Yukawa matrices first, and then 
introduce couplings to the vector-like states.

We represent the three generation of standard model lepton doublets by ${L_i}_L$ and the right-handed charged leptons by ${E_i}_R$, for $i=1, \ldots,  3$,  
We imagine that the Yukawa couplings are determined by a flavor symmetry of the form $Z_N \times G_F$.   Our interest is in the effect of the $Z_N$ factor, 
while we do not commit to any specific $G_F$.  We aim to show that the restrictions that follow from the $Z_N$ symmetry are sufficient to suppress the flavor 
mixing effects that we would like to avoid, while remaining compatible with a variety of possible flavor models that may determine the remaining, detailed 
structure of the Yukawa matrices.

We represent an element of $Z_N$ by $\omega^j$, for $j=1,\ldots N$, where $\omega^N \equiv 1$.   We assign the following transformation properties to 
the $L$ and $E$ fields, representing them here as column vectors:
\begin{equation}
L_L \rightarrow \Omega\, L_L \,\,\,\,\, \mbox{ and } \,\,\,\,\, E_R \rightarrow \Omega\, E_R \,\,\, ,
\end{equation}
where 
\begin{equation}
\Omega = \left(\begin{array}{ccc} 1 & 0 & 0 \\ 0 & \,\, \omega^{-n} & 0 \\ 0 & 0 & \,\, \omega^{n} \end{array}\right) \,\,\, .
\end{equation}
Note that $\omega^{-n} \equiv \omega^{N-n}$.  Assuming that the standard model Higgs doublet is unaffected by the $Z_N$ symmetry, the transformation properties of the charged-lepton
Yukawa matrix entries that lead to invariant couplings are summarized by
\begin{equation}
Y_E \sim \left(\begin{array}{c|cc} 1 & \omega^n & \omega^{-n} \\ \hline \omega^{-n} & 1 & \omega^{-2n} \\ \omega^n & \omega^{2n} & 1 \end{array}\right) \,\,\, ,
\end{equation}
where the transformation property of, for example, the $1$-$2$ entry is understood to be $Y_E^{12} \rightarrow   \omega^n \, Y_E^{12}$, and so on.   We will choose $N=2 \, n$ so that the entire two-by-two block on the lower right is 
unconstrained by the $Z_N$ symmetry, the least restrictive possibility that meets our needs\footnote{This choice is also compatible with $G_F$ having a non-Abelian
component in which two flavors of standard model leptons transform as a doublet.  However, it is sufficient (and simplest) for present purposes to imagine that $
G_F$ has only Abelian factors.}.  The amount by which the electron mass eigenstate is affected by the second and third generation fields, however, is entirely controlled by the size of $n$, once $Z_N$ breaking fields are introduced, as we discuss later.

A symmetry affecting the left-handed charged leptons also affects the left-handed neutrinos, so we must verify that neutrino phenomenology is not adversely 
affected.  For example, if we had imposed a $Z_2$ symmetry, with $n=1$, and required it to remain exactly unbroken, we can also completely eliminate mixing 
between the first generation charged leptons and those of the second and third generations.  However, if we then introduce three generations of right-handed 
neutrinos $N_i$, for $i=1, \ldots, 3$, one can show that there are no $Z_2$ charge assignments for the $N$ fields that leads to the correct neutrino mass squared 
differences and mixing angles, assuming the light mass eigenstates follow from the see-saw mechanism.   However, more favorable results may be obtained when the $Z_N$ symmetry is spontaneously broken.  Here, we assume the same 
transformation for all three right-handed
neutrino fields:
\begin{equation}
N_R \rightarrow \omega^p \, N_R \,\,\, ,
\end{equation}
where $p$ is an integer.  Defining the Dirac neutrino mass via ${\cal L} \supset \overline{L}_L \widetilde{H} Y_{LR} N_R + \mbox{h.c.}$, the transformation properties of the Yukawa coupling is given by
\begin{equation}
Y_{LR} \sim \left(\begin{array}{c|cc} \omega^{-p} & \omega^{-p} & \omega^{-p} \\ \hline \omega^{-n-p} & \omega^{-n-p} & \omega^{-n-p} \\ \omega^{n-p} & \omega^{n-p} & \omega^{n-p} \end{array}\right) \,\,\, .
\end{equation}
For the choice $n = 2 \, p$, or equivalently $N = 2 \, n = 4 \, p$, we may use the fact that $\omega^{-n-p} \equiv \omega^p$ and $\omega^{n-p} \equiv \omega^p$ to write
\begin{equation}
Y_{LR} \sim \left(\begin{array}{c|cc} \omega^{-p} & \omega^{-p} & \omega^{-p} \\ \hline \omega^{p} & \omega^{p} & \omega^{p} \\ \omega^{p} & \omega^{p} & \omega^{p} \end{array}\right) \,\,\, .
\end{equation}
The significance of this form is clear if we assume that there is a flavon field $\rho$ with the
$Z_N$ transformation property
\begin{equation}
\rho \rightarrow \omega \, \rho \,\,\, ,
\end{equation}
and a vacuum expectation value such that $\langle \rho \rangle/M \equiv \epsilon $ is a small parameter. Here $M$ is the flavor scale, which is the ultraviolet cut off of the effective theory. Then all the entries of $Y_{LR}$ are 
non-vanishing, and proportional to either $(\rho/M)^p$ or to $(\rho^*/M)^p$.   Hence, we may write
\begin{equation}
Y_{LR} = \epsilon^p \, \widetilde{Y}_{LR} \,\,\, ,
\label{eq:lrt}
\end{equation}
where $\widetilde{Y}_{LR}$ is a three-by-three matrix that is thus far arbitrary.  Following a similar argument, we define the right-handed neutrino Majorana mass 
matrix by the Lagrangian term $\overline{N^c}_R M_{RR} N_R$, and see immediately that
\begin{equation}
M_{RR} \rightarrow \omega^{-2p} M_{RR} \,\,\, .
\end{equation}
Again, this is consistent with the transformation property of $(\rho^*/M)^{2p}$, so we may write
\begin{equation}
M_{RR} = \epsilon^{2p} \widetilde{M}_{RR} \,\,\, ,
\label{eq:rrt}
\end{equation}
where $\widetilde{M}_{RR}$ is a three-by-three Majorana mass matrix that is also arbitrary thus far.  With $\widetilde{Y}_{LR}$ and 
$\widetilde{M}_{RR}$ arbitrary, it is possible to obtain any desired neutrino phenomenology, which demonstrates that the $Z_N$ symmetry
does not lead to unwanted phenomenological restrictions.  Theories that predict the detailed structure of $\widetilde{Y}_{LR}$ and 
$\widetilde{M}_{RR}$ by the breaking of an additional symmetry $G_F$ are compatible with this framework. Note that the overall powers of $\epsilon$
in Eqs.~(\ref{eq:lrt}) and (\ref{eq:rrt}) scale out of the see-saw formula which determines the Majorana
mass matrix for the three light neutrino mass eigenstates
\begin{equation}
M_{LL} = M_{LR} \, M_{RR}^{-1} \, M_{LR}^T ,\,\,\, 
\end{equation}
where $M_{LR} = (v/\sqrt{2}) \, Y_{LR}$.
The effect of the $Z_N$ symmetry on the form of the charged lepton Yukawa matrix is to
impose the form
\begin{equation}
Y_E \sim \left(\begin{array}{c|cc} y_{11} & \epsilon^n \tilde{y}_{12} & \epsilon^n \tilde{y}_{13} \\ \hline \epsilon^n \tilde{y}_{21} & y_{22} & y_{23} \\ \epsilon^n \tilde{y}_{31} & y_{32} & y_{33} \end{array}\right) \,\,\, .
\end{equation}
For $\epsilon$ sufficiently small, or $n$ sufficiently large, or both, we can make $Y_E$ as close to block diagonal as we like.

Now we include the vector-like state $\chi$ with the same electroweak quantum numbers as a right-handed electron, but charged also under a dark gauge group. Yukawa couplings involving 
$\overline{\chi}_L e_R$ and a dark Higgs field are unaffected by the $Z_N$ symmetry, while those involving $\overline{\chi}_L \mu_R$  or $\overline{\chi}_L \tau_R$ transform by $\omega^{\pm n}$.   
These potential sources of unwanted mixing that may emerge after the dark gauge symmetry is spontaneously broken are therefore highly suppressed by the same factors of $\epsilon^n$ that 
appear in the unwanted entries in $Y_E$.      We conclude that it is possible to make the $\chi$, $e$, $\mu$, $\tau$ mass matrix as block diagonal as desired, 
by suitable choice of $\epsilon^n$,  such that $\chi$ mixes substantially only with $e$, or any one desired lepton flavor, by a similar construction. 

The question of which lepton flavor is selected to mix with the heavier, vector-like states impacts the phenomenology of the dark gauge bosons.  For example, if the mixing only involves the
$\tau$ lepton, then bounds on the $A_D^a$ from searches for $s$-channel resonances in low-energy $e^+ e^-$ collisions, or from indirect processes like the electron or muon $g-2$
would be irrelevant.  The phenomenology in the case where the mixing involves either a first or second generation lepton would lead to more meaningful constraints, but one that would depend on
other assumptions about the spectrum, for example if $A_D^3$ decays visibly or invisibly, which depends on the dark matter mass.   In the following section, we will assume the least constrained possibility, that the $\chi$'s mix with the $\tau$, and consider the wide range of phenomenological issues associated with the other two possibilities in separate 
work~\cite{shikha}.   This has the appealing aesthetic feature that the flavor symmetry distinguishes the third generation from the other two, an idea that has appeared in many other contexts in
the literature on the flavor structure of the standard model~\cite{some}.

\section{Phenomenology}  \label{sec:pheno}

\subsection{Relic Density}

The scattering amplitude for $s$-channel dark matter annihilation into standard model particles depicted in Fig.~\ref{fig:one}, with $e$ replaced by $\tau$,  
is given by
\beq
\mathcal M( \psi^{(1)} \overline{\psi^{(1)}} \to \tau^+ \tau^-) 
=\frac{ ig_D^2\theta^2}{4\big(q^2-m_{A_D^3}^2+im_{A_D^3}\Gamma^D\big)} \,\, \overline v(p') \gamma^\mu u(p) \,\, \overline u(k) \gamma_\mu v(k')
\enq
where $p\; (p')$ is the momentum of the incoming dark matter fermion (anti-fermion), $k\; (k')$ is the the momentum of the outgoing $\tau^-$ 
($\tau^+$) and $q=p+p'$ is the momentum flowing through the $A_D^3$ propagator. As discussed in Sec.~\ref{sec:model},
the lightest gauge boson $A_D^3$ couples to the vector-like states $\chi^{(1)}$ and $\chi^{(2)}$, which then mix with a standard model 
lepton flavor (chosen here as $\tau$) after spontaneous symmetry breaking. This results in a factor of $\theta^2$, defined in 
Eq.~(\ref{eq:ourtheta}), in the scattering amplitude. 

Our numerical results for dark matter annihilation depend on assumptions about the dark particle mass spectrum and couplings.  We assume the picture described earlier, where the lightest states consist of $\psi^{(1)}$ and $A^3_D$, and decays of  $A^3_D$ 
to any of the heavier exotic states are not kinematically allowed.  For the mass range studied in this section, $A_D^3$ can decay to $\tau^+\tau^-$, and possibly also 
$\psi^{(1)}\overline{\psi^{(1)}}$, depending on the dark matter mass.  Consequently, the total decay width of the dark gauge boson 
appearing in the propagator is given by
\begin{equation}
\Gamma^D = \Gamma\left(A_D^3 \to \tau^+\tau^-\right) + \Theta \left( m_{A_D^3} - 2 \, m_{\psi^{(1)}}\right)\, \Gamma\left(A_D^3 \to \psi^{(1)}\overline{\psi^{(1)}}\right)
\end{equation}
where $\Theta$ is a step function, {\em i.e.}, $\Theta(x)=1 \text{ if } x\geq0 \text{ and }  \Theta(x)=0 \text{ if } x<0$, and
\begin{equation}
\Gamma\left(A_D^3 \to \tau^+\tau^-\right) = \frac{1}{48\pi} g_D^2 m_{A_D^3} \theta^4 \left(1+\frac{2\, m_\tau^2}{m_{A_D^3}^2}\right)\left(1-\frac{4\, m_\tau^2}{m_{A_D^3}^2}\right)^{1/2} \,\,\, ,
\end{equation}
\begin{equation}
\Gamma\left(A_D^3 \to \psi^{(1)}\overline{\psi^{(1)}}\right) =  \frac{1}{48\pi} g_D^2 m_{A_D^3}  \left(1+\frac{2\, m_{\psi^{(1)}}^2}{m_{A_D^3}^2}\right)\left(1-\frac{4\, m_{\psi^{(1)}}^2}{m_{A_D^3}^2}\right)^{1/2} \,\,\, .
\end{equation}
Since the mean dark matter velocity is typically around 220 km/s~\cite{pdg}, we work in the non-relativistic limit where
$E_{\psi^{(1)}} \approx m_{\psi^{(1)}}$.  We then find the thermally averaged annihilation cross section times velocity
 \beq
 \label{eq:crosssection}
 \braket{\sigma_A v} =  \frac{g_D^4 \theta^4}{32\pi}\frac{2m_{\psi^{(1)}}^2+m_\tau^2}{(4m_{\psi^{(1)}}^2-m_{A_D^3}^2)^2+m_{A_D^3}^2 \Gamma_D^2}  \left(1-\frac{m_\tau^2}{m_{\psi^{(1)}}^2}\right)^{1/2}. 
 \enq
Using this we calculate the freeze-out temperature $T_F$ and the dark matter relic density by standard methods \cite{Kolb:1990vq}. Dark matter freeze out occurs when the interaction probability per unit time $\Gamma_{\psi^{(1)}}$, equals the 
expansion rate of the universe, $H$, {\em i.e.}, 
\beq
\label{eq:freezeout}
 \left.\frac{\Gamma_{\psi^{(1)}}}{H} \right|_{T=T_F} = \left.\frac{n_{EQ}^{\psi^{(1)}} \braket{\sigma_A v}}{H}\right|_{T=T_F} \simeq 1.
 \enq
 Here $n_{EQ}^{\psi^{(1)}}$ is the equilibrium number density of the dark matter particle, given by 
 \beq
 n_{EQ}^{\psi^{(1)}} = 2 \left( \frac{m_{\psi^{(1)}} T}{2\pi}\right)^{3/2} e^{-m_{\psi^{(1)}}/T}.
 \enq
Freeze-out occurs during the radiation-dominated epoch in which case
\beq
H = 1.66 \,g_*^{1/2} T^2 / M_{pl},
\enq
where $M_{pl} = 1.22 \x 10^{19}$ GeV is the Planck mass and $g_*(T)$ the number of relativistic degrees of freedom at temperature $T$, 
\beq
g_*(T) = \sum_{i = bosons} g_i \left(\frac{T_i}{T}\right)^4 + \frac{7}{8} \sum_{i=fermions} g_i \left(\frac{T_i}{T}\right)^4.
\enq
Finally the dark matter relic density is given by
\beq
\label{eq:relicdensity}
\Omega_D h^2 = \frac{2\cd(1.07 \x 10^9 \text{ GeV}^{-1})\, x_F}{\sqrt{g_*(T_F)} M_{Pl} \bk{\sigma_A v}}.
\enq
We define $x_F \equiv m_{\psi^{(1)}}/ T_F$ where $T_F$ is obtained by solving Eq.~(\ref{eq:freezeout}). The factor of 2 is included because we are accounting for the density of dark matter particles and antiparticles. We require Eq.~(\ref{eq:relicdensity}) to 
reproduce the WMAP result $0.1186 \pm 0.0020$ \cite{pdg} within two standard deviations.

To display our results, we fix $m_{A^3}$ and $\theta$ and find the regions of the $g_D$-$m_{\psi^{(1)}}$ plane in which the desired dark matter
relic density is obtained.  We assume that the mixing angle remains small ($\theta<1$) but not so small that a satisfactory dark matter annihilation cross section cannot be obtained.  So that the dark gauge coupling remains perturbative, we assume 
$\alpha_D / (4\pi) < 1/3$ or equivalently $g_D < 4\pi/\sqrt3 \sim 7.25$; one-loop corrections become comparable to tree-level amplitudes when $\alpha / (4\pi) \approx 1$, so one-third of this value is a reasonable upper limit on the dark coupling constant. For the purposes of 
determining $g_*$, we assume all exotic mass eigenstates other than $\psi^{(1)}$ and $A_D^3$, are at $m_Z= 91.1876$~GeV.   With this choice, the $Z$ boson cannot decay into $\chi \overline{\chi}$ or $\chi \tau$, which could lead to an unacceptable broadening of the precisely measured $Z$ boson width~\cite{pdg}.

\begin{figure}[h]
\centering
	\subfigure{%
		\includegraphics[scale=0.6]{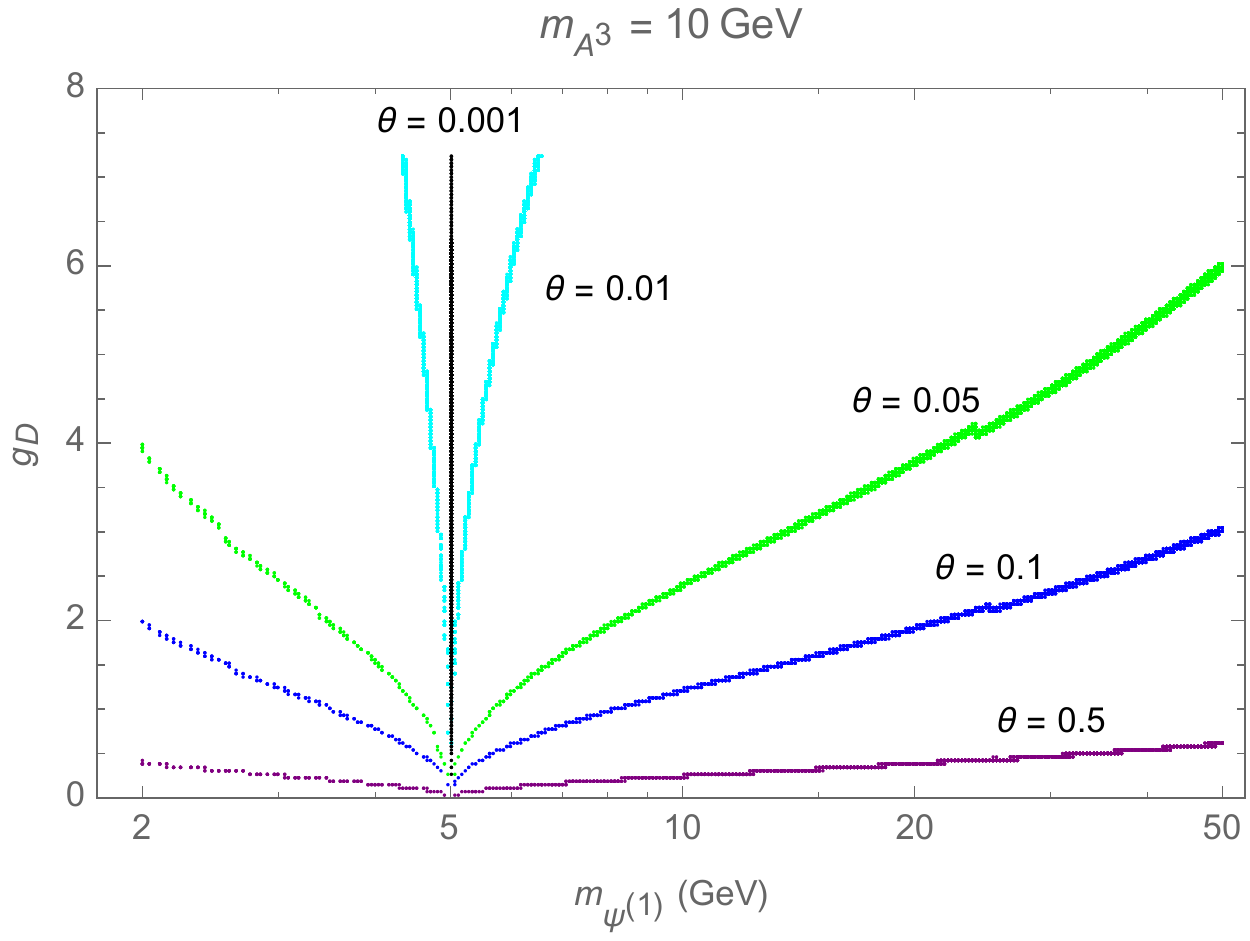}} \hspace{2em}%
	\subfigure{%
		\includegraphics[scale=0.6]{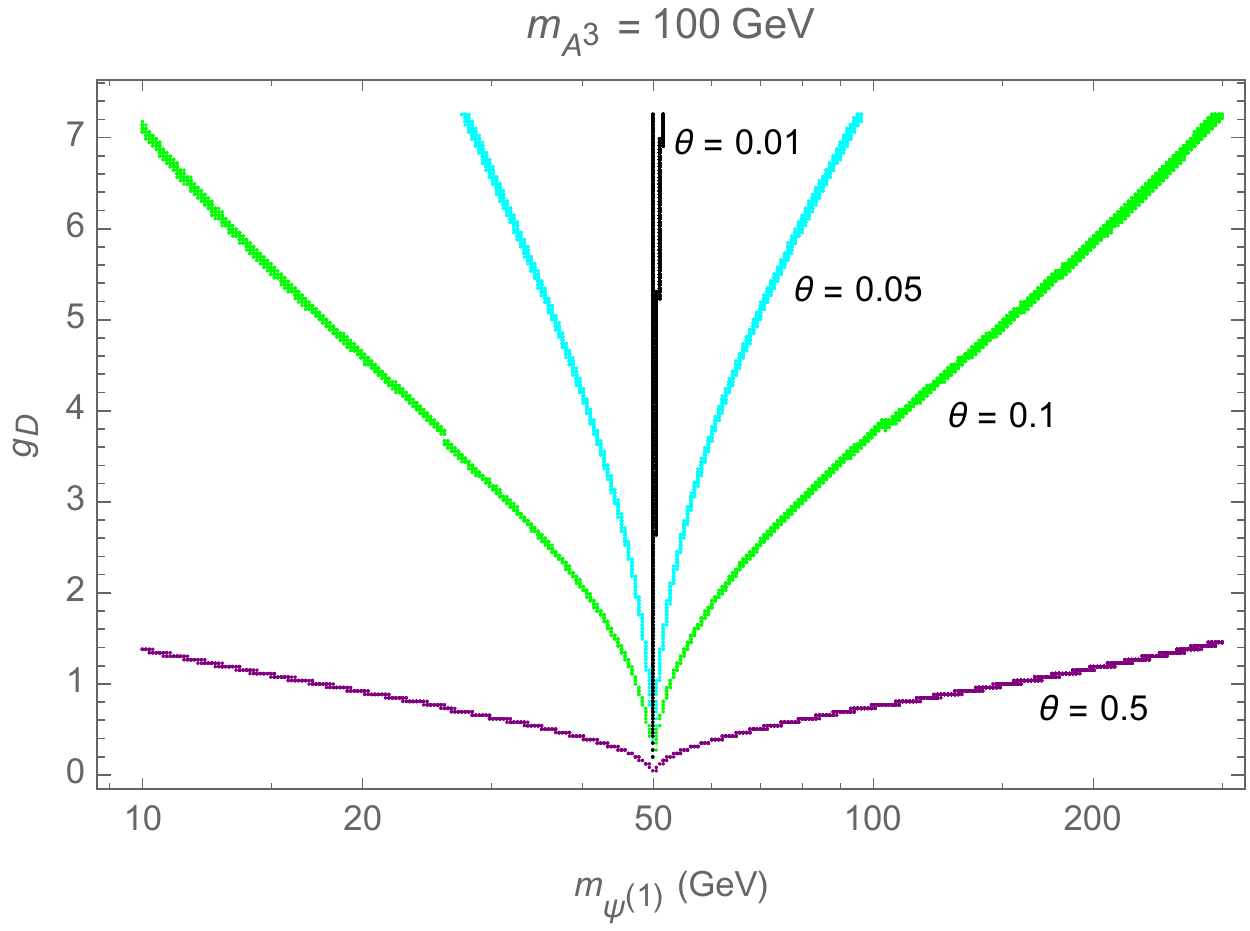}}%
	\caption{Regions of the $g_D$-$m_{\psi^{(1)}}$ plane in which the dark matter relic density is within two standard deviations of the WMAP result $0.1186 \pm 0.0020$~\cite{pdg}, for fixed choices of $m_{A^3}$ and $\theta$.  The allowed bands are not 
	perfectly smooth due to their dependence on $g_*$, which is not a continuous function.   The point of minimum $g_D$ corresponds to resonance annihilation, where $m_{\psi^{(1)}} = m_{A^3}/2$. Note that as $\theta$ decreases the range of $m_{\psi^{(1)}}$ in which $g_D$ remains 
	perturbative moves towards the resonance region. \label{fig:plots}}
\end{figure}

Fig.~\ref{fig:plots} shows the regions of the $g_D$-$m_{\psi^{(1)}}$ plane in which the dark matter relic density is within two standard deviations 
of the WMAP result $0.1186 \pm 0.0020$~\cite{pdg}, for fixed choices of $m_{A^3}$ and $\theta$.  We have intentionally centered the plots 
around the point of resonance annihilation $m_{\psi^{(1)}} = m_{A^3}/2$ where the cross section is largest.  For small values of $g_D$ at fixed $\theta$, some tuning is required to achieve a large enough annihilation cross section.  However, Fig.~\ref{fig:plots} 
indicates that we can have larger, perturbative values of $g_D$ without requiring that we sit unnaturally close to the resonance.  As $\theta$ is made progressively smaller, however, more tuning is required.  This is indicated by the narrowing 
range in $m_{\psi^{(1)}}$ for each solution in which $g_D$ is also perturbative.

Of course, the values of $\theta$ that are indicated in Fig~\ref{fig:plots}  are related to choices for the masses and coupling in the model, such that $\theta^2 = g_D \left(\frac{m_1^2}{M_1^2}-\frac{m_2^2}{M_2^2} \right)$, where the $m_i$ and
$M_i$ were defined in Eq.~(\ref{eq:bigM}).   It is not hard to verify that the values of $\theta$ shown in Fig.~\ref{fig:plots} can be achieved given the assumptions that went into the making of 
the plots. For example, in the $m_{A^3}=10$~GeV plot, consider the point where $g_D \approx 1$ and $m_{\psi^{(1)}} \approx 8.5$~GeV, on the $\theta=0.1$ band.   Given our earlier 
assumption in computing $g_*$ that the heavier exotic states are at $m_Z$, one can check that this is consistent with, for example, $v_{D1} = v_{D2} \approx 14$~GeV,  $v_T \approx 49$~GeV, $\lambda_s \approx 0.85$, and $y_1=y_2 \approx 0.06$, where the Yukawa couplings $y_i$ were defined in Eq.~(\ref{eq:lchi}).  Similar statements can be made about other points on the allowed bands\footnote{The scenario that we have considered assumes that communication between dark and visible sectors occurs primarily through the portal that we have proposed, involving mixing with vector-like leptons.   It is of course possible to have scenarios in which communication is also significant through Higgs portal couplings or other mediators.  The results presented in this section demonstrate the effectiveness of the portal we have proposed taken in 
isolation; this may nonetheless accurately represent a subset of the parameters space of a more complicated model with other dark matter annihilation channels.}.

\subsection{Direct Detection}

The interactions that we have discussed to this point have involved leptons exclusively, but couplings to quarks that are generated at the loop level also have significant consequences.  
In this section, we consider direct detection of the dark matter in the model via dark-matter-nucleon elastic scattering.   The couplings to quarks arise after the SU(2)$_D$ symmetry is 
spontaneously broken, since kinetic mixing between $A^3_D$ and hypercharge is then allowed, via an effective dimension-5 operator 
\begin{equation} 
\mathcal{L}_{eff} = X \tr\left(\langle H_{T}\rangle T^{a}A^{a}_{D \mu\nu}\right)Y^{\mu\nu} \,\,\, , \label{eq:Leff1} 
\end{equation}
\begin{figure}[t]
\centering
\includegraphics[width=0.75 \textwidth]{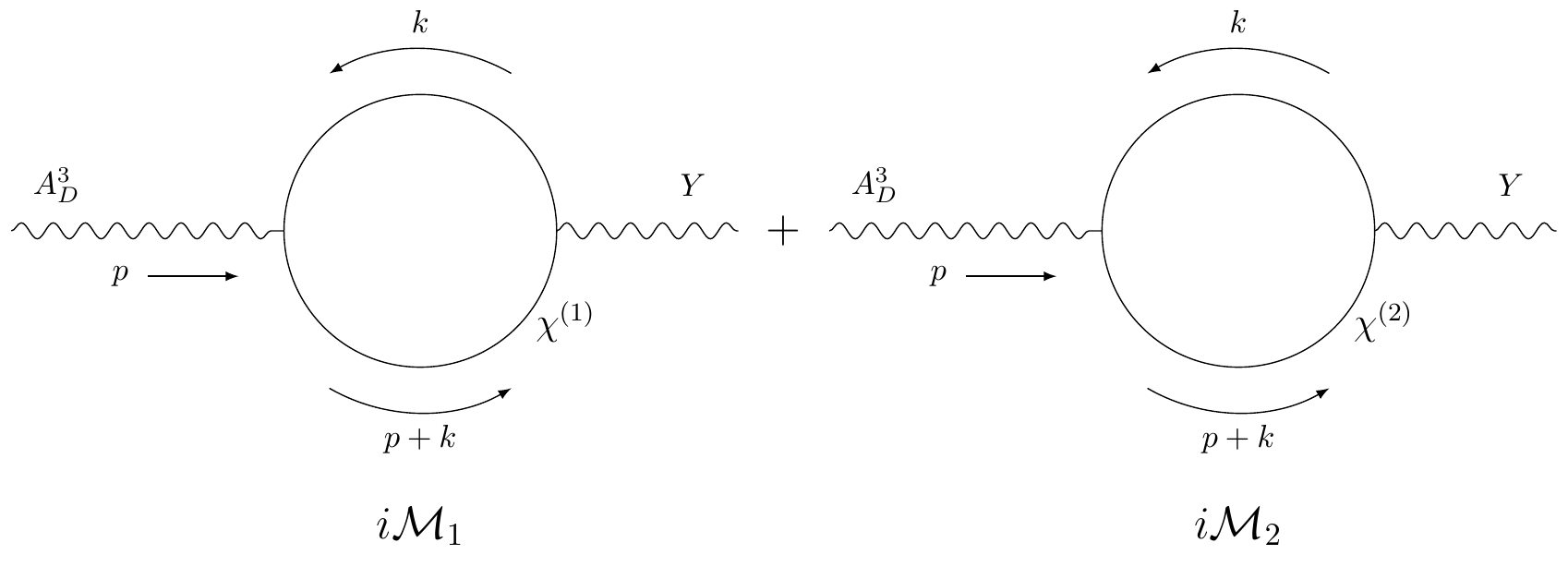}
\caption{Self energies leading to kinetic mixing between the third dark gauge boson $A_{D}^{(3)}$ and hypercharge $Y$ after SU(2)$_D$ is spontaneously broken. \label{fig:vertex}} 
\end{figure}	
where we have set the triplet Higgs to its vev, as per Eq.~(\ref{eq:hdhtv}).   Here, $X$ is a constant with units of GeV$^{-1}$ which is found by integrating out the ``heavy" physics, {\em i.e.},  the $\chi$ fields, the only fields that are charged both under SU(2)$_D$ and hypercharge U(1)$_Y$.   To proceed, we study the self-energy shown in Fig.~\ref{fig:vertex}, where 
$\chi^{(1)}$ and $\chi^{(2)}$ here represent the heavy mass eigenstates, whose mass eigenvalues are given approximately by $m_{\xone} = M_\chi -\delta$ and $m_{\xtwo} = M_\chi + \delta$ 
where $\delta \equiv \lambda_s' \, v_T/2$.  (For the purposes of this estimate, we ignore mass mixing with the standard model lepton, which is a subleading correction.)   The first diagram is 
given by
\begin{equation}  
		i\M_{1} =  - \frac{g_{D} g_{Y}}{2}  \int \frac{d^{4}k}{(2\pi)^{4}} 
		\frac{ 
			\tr\left[ \gamma^{\mu} \left( \slashed{k} + m_{\xone} \right) \gamma^{\nu} \left( \slashed{k} + \slashed{p} +m_{\xone} \right) \right] }{ [ k^{2}-m_{\xone}^{2}+i\epsilon ] [ (k+p)^{2}-m_{\xone}^{2}+i\epsilon] } \, \, \, . 
\end{equation} 
After carrying out this loop integral using dimensional regularization in $D=4-\epsilon$ dimensions, the amplitude is 
\begin{equation} 
i\mathcal{M}_{1} =- \frac{g_{D}g_{Y}}{8\pi^{2}} \int_{0}^{1} dx \ x(1-x) \left(\frac{4}{\epsilon}-2\gamma +2\log(4\pi) -2 \log(\Delta_{1})\right) i (g^{\mu\nu}p^{2}-p^{\mu}p^{\nu}) \, \, \, , 
\end{equation}
where $\Delta_{1} = m_{\xone}^{2} -x(1-x)p^{2}$.   Since $A_D^3$ couples to the $\chi$ proportional to $\sigma^3/2$, the amplitude $i\M_{2}$ shown in 
Fig.~\ref{fig:vertex} will differ from $i\M_{1}$ by a overall minus sign and the replacement of the $\chi^{(1)}$ by the $\chi^{(2)}$ mass.  Hence, $\Delta_{1}$ is replaced by $\Delta_{2} = m_{\xtwo}^{2} -x(1-x)p^{2}$. Then, when these two amplitudes are added together, all terms in the remaining Feynman parameter integral cancel, except for the terms that depend on the fermion masses: 
\begin{equation} 
i\mathcal{M}_{1} + i \mathcal{M}_{2} = i(g^{\mu\nu}p^{2} - p^{\mu}p^{\nu}) \frac{g_{D}g_{Y}}{4\pi^{2}} \int_{0}^{1}dx\ x(1-x) \log\left(\frac{\Delta_{1}}{\Delta_{2}}\right) \,\,\, . 
\end{equation}
Assuming the mass splitting $\delta$ is small compared to the $\chi$ masses (which will turn out to be the case) the integrand can be expanded in $\delta$.  The leading order term can be found
using $x(1-x)\log(\Delta_{1}/\Delta_{2}) \approx -\frac{4mx(1-x)}{m^{2}-x(1-x)p^{2}} \delta$.  Moreover, we can also expand the result in powers of momentum, which can later be compared to a 
derivative expansion in the low-energy effective theory.  We find
\begin{equation} 
i\mathcal{M}_{1} + i \mathcal{M}_{2} = - i \frac{g_{D}g_{Y}\delta}{6\pi^{2} M_\chi}(g^{\mu\nu}p^{2} - p^{\mu}p^{\nu}) + \cdots \, \, \, , \label{eq:chiloop}
\end{equation}
where the $\cdots$ represents terms involving higher powers of $\delta$ and $p^2/M_\chi^2$.   The result in Eq.~(\ref{eq:chiloop}) must be matched to a similar amplitude in the low-energy effective theory in which the $\chi$ fields have been integrated out.  We identify this as the tree-level amplitude associated with the Eq.~(\ref{eq:Leff1}), treated as a two-point vertex,
\begin{equation} 
i\mathcal{A}  = i X v_{T} \left( p^{2}g^{\mu\nu}-p^{\mu}p^{\nu} \right) \,\,\, , \label{eq:AmpEff} 
\end{equation} 
from which we conclude
\begin{equation}
X = - \frac{g_{D}g_{Y}\delta}{6\pi^{2} M_\chi v_{T}} \, \, \, . \label{eq:KineticMixing}
\end{equation}
\begin{figure}
\centering
\includegraphics{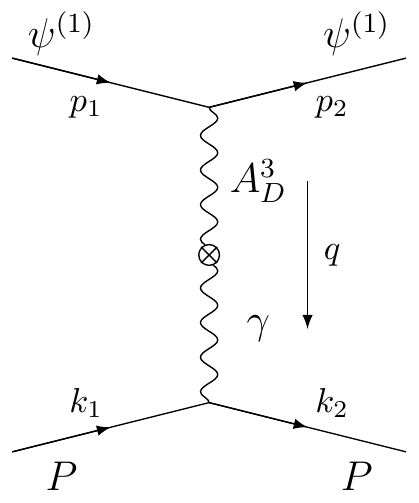}
\caption{The Feynman diagram for the scattering of the dark matter particles, $\psi^{(1)}$, off of protons, $P$, through kinetic mixing of the dark matter boson $A_{D}^{3}$ and the photon, $\gamma$.  \label{fig:crosssection}} 
\end{figure}

Using Eqs.~(\ref{eq:Leff1}) and (\ref{eq:KineticMixing}), we can now calculate the cross section for dark matter scattering off of nucleons. We will be working in the limit of low momentum transfer 
$q \sim {\cal O}(100)$~MeV ($\ll M_\chi$), where the effective description is accurate and where scattering through the $Z$ boson is suppressed by $q^2/m_Z^2 \sim 10^{-6}$ compared to the photon.  Hence,
we will consider kinetic mixing involving the photon only from here on. First, we consider the dark matter, $\psi^{(1)}$, scattering off of a quark, $q_{f}$, as in the diagram in Fig.~\ref{fig:crosssection}, with the protons replaced by a quark of flavor $f$.   This can be described by the effective dimension-six operator
\begin{equation}
 \mathcal{L}_{eff,q} = C_{f} \, \overline{\psi^{(1)}} \gamma^{\mu}\psi^{(1)} \overline{q_{f}}\gamma_{\mu}q_{f} \, \, \, . \label{eq:effqop}
 \end{equation} 
In the full theory, this quark-dark matter scattering amplitude is 
\begin{equation} 
i\M_{f} = i X v_{T} Q_{f} \frac{g_{D}}{2} e \frac{1}{(q^{2}-m_{A^{3}_{D}}^{2} +i\epsilon)(q^{2}+i\epsilon)} \overline{\psi^{(1)}}\gamma^{\mu}\psi^{(1)}\overline{q_{f}}\gamma_{\mu}q_{f} 
\end{equation} 
or, in the limit of $q^{2}\ll m_{A^{(3)}_{D}}^{2}$, 
\begin{equation} 
i\M_{f}  =-i \frac{Xv_{T}Q_{f}}{m_{A^{3}_{D}}^{2}} \frac{g_{D}}{2} e \overline{\psi^{(1)}}\gamma^{\mu}\psi^{(1)}\overline{q_{f}}\gamma_{\mu}q_{f} \, \, \, . 
\end{equation}
From this, we conclude the coefficient $C_{f}$ for quarks is 
\begin{equation}
C_{f} = - \frac{g_d e Xv_{T}Q_{f}}{2 m_{A_{D}^{3}}^{2}} = \frac{g^2_{D} e^2 \delta Q_{f}}{12 \pi^{2} M_\chi m_{A_{D}^{3}}^{2} } \,\,\, . 
\end{equation} 

Of interest, however, is the effective interactions involving nucleons rather than quarks, which can be written
\begin{equation}
\mathcal{L}_{eff,N}  = C_n \, \overline{\psi^{(1)}}\gamma^{\mu} \psi^{(1)} \overline{n} \gamma_{\mu}n + C_p \, \overline{\psi^{(1)}}\gamma^{\mu} \psi^{(1)} \overline{p} \gamma_{\mu} p  \,\,\,.
\label{eq:effNop}
\end{equation}
Using the fact that the quark vector currents are conserved, so that the spatial integral of the zeroth component is a quark number operator, one can match matrix elements of Eq.~(\ref{eq:effqop})
between nucleon states with the same for Eq.~(\ref{eq:effNop}), from which one concludes $C_{n} = C_{u}+ 2 C_{d}$ and $C_{p} = 2 C_{u} + C_{d}$, for the neutron and proton, respectively.  (There 
are no form factors as there would be for scalar quark operators.)   Since the flavor dependence of the $C_f$ comes only from the electric charge, the coefficient $C_n$ and thus the scattering amplitude for $\psi^{(1)}$ off of neutrons are both zero. Therefore, the only relevant scattering is with the proton, for which
\begin{equation} 
C_{p} = 2\, C_{u}+C_{d} = \frac{g^2_{D} e^2 \delta }{12 \pi^{2} M_\chi m_{A_{D}^{3}}^{2} }\, \, \, . 
\end{equation} 
Taking into account that the dark matter is non-relativistic and that momentum transfers are small, a straightforward calculation of the scattering cross section yields		
\begin{equation} 
\lara{\sigma_{\psi^{(1)}p\rightarrow\psi^{(1)}p}} = \frac{g_{D}^{4}e^{4}m_{p}^{2}m_{\psi^{(1)}}^{2} }{ 576 \pi^{5} \left( m_{p} + m_{\psi^{(1)}} \right)^{2} m_{A^{3}_{D}}^{4} } 
\left( \frac{2\delta}{M_\chi}\right)^{2} \,\,\, , \label{eq:crosssection2} 
\end{equation}
where we have separated out the dependence on $2\delta/M_\chi$, the fractional mass spitting of the vector-like leptons.  Since this splitting is a free parameter in our model, we can use the
experimental bounds on the dark-matter-nucleon elastic scattering cross section to say something about the vector-like lepton spectrum.

\begin{figure}
\centering
\includegraphics[width=.4\textwidth]{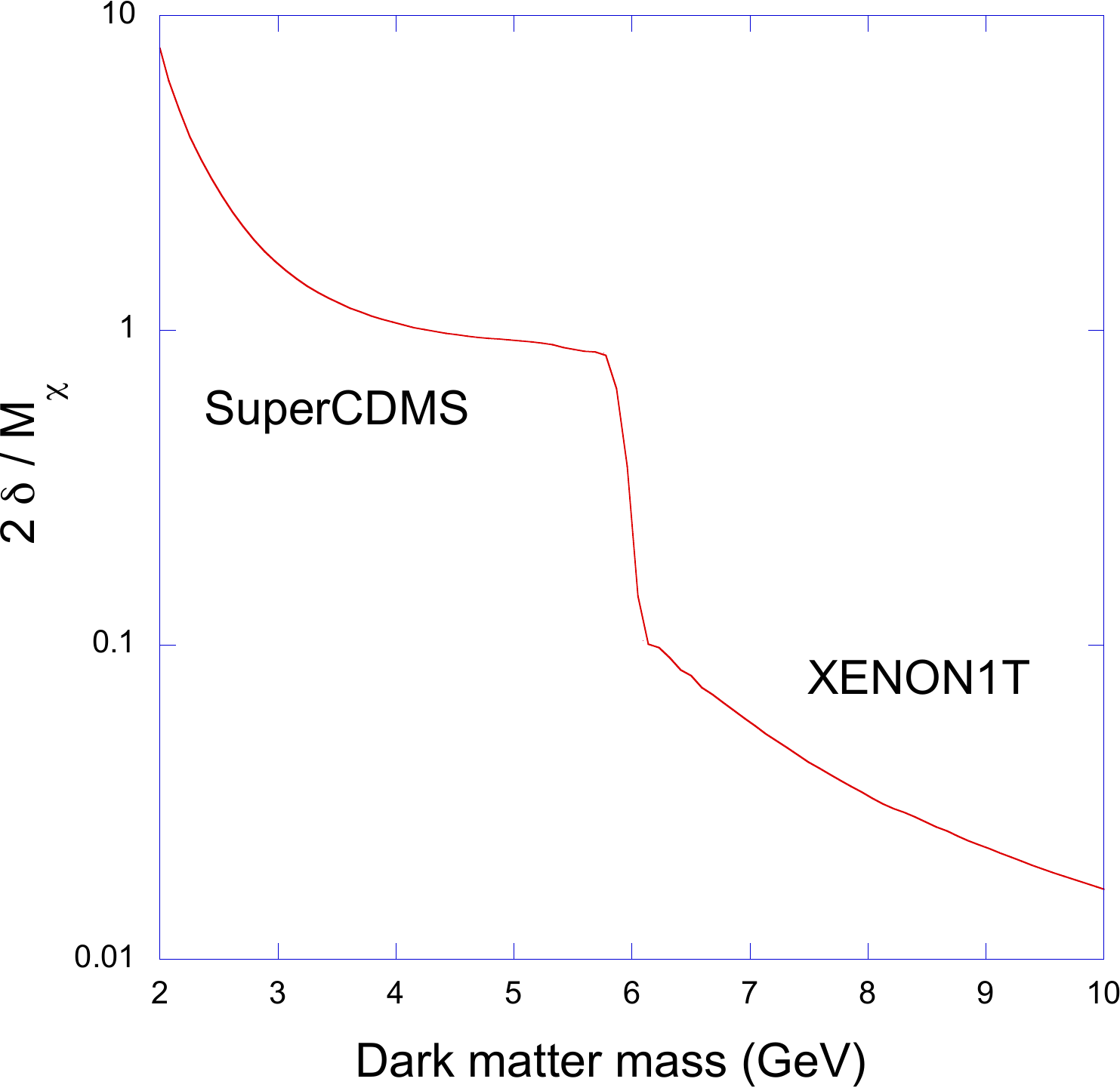}
\caption{Upper bound on the fractional mass splitting of the $\xone$ and $\xtwo$ fermions as a function of the mass of the dark matter particle, $\psi^{(1)}$, assuming  $g_{D}=0.3$ and  
$m_{A^{3}_{D}} = 10\, \mbox{GeV}$.  The discontinuity in the curve reflects that the bounds on the dark-matter-nucleon elastic scattering cross section originate from the CDMSlite experiment~\cite{CDMSlite}
below $m_{\psi^{(1)}} \approx 6$~GeV, where the otherwise tighter bounds from the XENON1T experiment~\cite{xenon1t} do not exist. \label{fig:directdetection}} 
\end{figure}

Using experimental bounds on the cross section from XENON1T~\cite{xenon1t} and CDMSlite~\cite{CDMSlite}, we show bounds on the $\chi^{(1)}$-$\chi^{(2)}$
mass splitting for dark matter masses between $2$~GeV and $10$~GeV.  The results of this calculation are shown in Fig.~\ref{fig:directdetection}, where a dark coupling of $g_{D}=0.3$ and a dark 
boson mass of $m_{A^{3}_{D}} = 10\, \mbox{GeV}$ have been used. For dark matter masses below approximately $6$~GeV, the cross section bounds from CDMSlite are used, since no data from XENON1T is
available in this region. Although there is CDMSlite data for dark matter masses above $6$~GeV, these bounds are superceded by the stricter ones from XENON1T.  For the  
range of $\psi^{(1)}$ masses in Fig.~\ref{fig:directdetection} that are affected by the XENON1T bounds, the masses of the charged fermions $\xone$ and $\xtwo$ are 
degenerate at the $1$-$10$\% level at minimum.  This feature could be observed in collider searches for the vector-like leptons and possibly correlated with a dark matter direct detection signal.

\section{Conclusions}
We have presented an explicit, renormalizable non-Abelian dark SU(2) model which contains two vector-like fermion doublets. One of them, $\psi$, includes a dark matter 
candidate; the other doublet, $\chi$, has the same electroweak quantum numbers as a right-handed electron, so that communication with the visible sector can occur via mass 
mixing. The $\psi$ and $\chi$ fields communicate with each other via the dark gauge group, so that the dark matter may annihilate to standard model leptons.  The dark SU(2) 
symmetry is spontaneously broken via a Higgs sector involving doublet and triplet fields.  The doublet vacuum expectation value (vev) leads to mixing between the $\chi$ and standard model lepton fields, while the triplet vev splits the mass spectrum leaving a simple lower-energy theory consisting of the dark matter (the lightest $\psi$ mass eigenstate) and
the mediator (the third component of the SU(2) gauge multiplet).  We identify a discrete flavor 
symmetry that allows mixing  between the vector-like leptons $\chi$ and a single standard model lepton flavor exclusively; the remaining standard model lepton flavors may 
mix only with each other.  This flavor sequestering eliminates lepton-flavor-violating effects, relaxing bounds on the vector-like lepton mass scale.  As a consequence, mixing
between the chosen lepton flavor and the $\chi$ can be large enough so that the correct relic density can be obtained exclusively via dark matter annihilation to lepton-anti-lepton pairs, for
perturbative values of the dark gauge coupling.  This is true even if no other significant annihilation channels are available that originate from other portals.

The structure of our model avoids complications that would ensue if we tried to couple the dark gauge bosons directly to standard model fields, such as the necessity of 
including extraneous fermions to cancel chiral anomalies, or special Higgs representations to allow for acceptable standard model Yukawa couplings.  Unlike some of the non-Abelian dark 
matter models appearing in the literature, the portal we present is renormalizable and completely specified, including the discrete flavor symmetries that control the pattern of mixing between 
exotic and standard model fermions states.   The portal we define for communication between the dark and visible sectors presents a well defined framework for answering 
phenomenological questions.  In the present work, we showed that there are regions of the dark gauge coupling - dark matter mass plane where the correct relic density is 
obtained, and where current direct detection bounds are satisfied.   The latter consideration also allowed us to conclude that the two heavy lepton mass eigenstates (roughly the 
two components of the $\chi$ doublet) are notably degenerate in mass (to keep kinetic mixing effects small), a feature that could be tested in collider searches for these states.   
This observation, together with the distinct lepton flavor structure of the $\chi$ decays, suggests that the collider signatures of the portal that we have proposed are 
worthy of future detailed investigation.
  
\begin{acknowledgments}  
This work was supported by the NSF under Grant PHY-1519644.
\end{acknowledgments}


\end{document}